\documentclass[runningheads]{llncs}
\usepackage{graphicx,wrapfig,lipsum,booktabs}
\usepackage{amsmath}
\usepackage{amssymb}
\usepackage{booktabs}
\usepackage{csquotes}
\usepackage{multirow}
\usepackage{mathtools}
\usepackage{makecell}
\usepackage[normalem]{ulem}
\usepackage[pagebackref,breaklinks,colorlinks]{hyperref}
\usepackage[usenames,dvipsnames,svgnames,table]{xcolor}
\usepackage{subcaption}
\usepackage{xspace}
\usepackage{amssymb, amsmath}
\usepackage{dsfont}
\usepackage{xcolor}
\makeatletter
\DeclareRobustCommand\onedot{\futurelet\@let@token\@onedot}
\def\@onedot{\ifx\@let@token.\else.\null\fi\xspace}
\def\eg{\emph{e.g}\onedot} 

\def\ie{\emph{i.e}\onedot}

\def\etal{\emph{et al}\onedot}
\makeatother
\DeclareMathOperator{\Loss}{\mathcal{L}}
\DeclareMathOperator*{\R}{\mathbb{R}}

\renewcommand{\vec}[1]{{\mathbf #1}}
\usepackage{pifont}
\usepackage{hyperref}
\hypersetup{
    colorlinks=true,
    linkcolor=blue,
    filecolor=magenta,      
    urlcolor=cyan,
}
\usepackage[font=scriptsize]{caption}
\usepackage{mwe}
\usepackage[capitalize]{cleveref}
\crefname{section}{Sec.}{Secs.}
\Crefname{section}{Section}{Sections}
\Crefname{table}{Table}{Tables}

\begin{document}
\title{Hybrid Window Attention Based Transformer Architecture for Brain Tumor Segmentation}
\titlerunning{CR-Swin2-VT}
%
\author{Himashi Peiris 
\inst{1}, Munawar Hayat 
\inst{3}, Zhaolin Chen 
\inst{1,2}, Gary Egan 
\inst{2}, Mehrtash Harandi 
\inst{1}}
\authorrunning{H. Peiris et al.}
%
\institute{Department of Electrical and Computer Systems Engineering, Monash University, Melbourne, Australia. \and Monash Biomedical Imaging (MBI), Monash University, Melbourne, Australia. \and Department of Data Science \& AI, Faculty of IT, Monash University, Melbourne, Australia. \\
\email{\{Edirisinghe.Peiris, Munawar.Hayat, Zhaolin.Chen, Gary.Egan, Mehrtash.Harandi\}@monash.edu}}
\maketitle              
%

%
%

\begin{abstract}
As intensities of MRI volumes are inconsistent across institutes, it is essential to extract universal features of multi-modal MRIs to precisely segment brain tumors. In this concept, we propose a volumetric vision transformer that follows two windowing strategies in attention for extracting fine features and local distributional smoothness (LDS) during model training inspired by virtual adversarial training (VAT) to make the model robust. We trained and evaluated network architecture on the FeTS Challenge 2022 dataset. Our performance on the online validation dataset is as follows: Dice Similarity Score of 81.71\%, 91.38\% and 85.40\%; Hausdorff Distance (95\%) of 14.81 mm, 3.93 mm, 11.18 mm for the enhancing tumor, whole tumor, and tumor core, respectively. Overall, the experimental results verify our method's effectiveness by yielding better performance in segmentation accuracy for each tumor sub-region. \href{https://github.com/himashi92/vizviva_fets_2022}{Our code implementation is publicly available}. 

\end{abstract}
\section{Introduction}
\label{sec:intro}
Interpreting clinically acquired, multi-institutional multi-parametric magnetic resonance imaging (mpMRI) scans is a long-standing challenge in the medical AI as these medical volumes consist of intrinsically heterogeneous lesions, tumors, or anatomical objects. 
Accurate segmentation is a prerequisite for clinical diagnosis and treatment planning.
Federated Tumor Segmentation (FeTs) challenge has clinically acquired mpMRI, and the task of the challenge is to segment intrinsically heterogeneous brain tumors (gliomas), and the objective is to create a consensus segmentation model acquired from various institutions without pooling their data together~\cite{pati2021federated,reina2021openfl,baid2021rsna,sheller2020federated}. Segmenting brain tumors from the medical scans is a tedious process and it requires expertise and concentration since tumors are heterogeneous. 
Therefore, implementing a generalized deep learning model that can produce reliable predictions or segmentation masks with precise regions of interest for patient data from various institutional distributions is highly demanded in scientific research. Traditionally, automated segmentation tools were implemented using manual feature engineering based learning methods such as decision forest~\cite{zikic2012decision}, conditional random field (CRF)~\cite{wu2014brain}. However, with the recent progress in Convolutional Neural networks (CNN) and improvements in computational resources (\eg, Graphical Processing Units (GPU)), deep learning models for tumor segmentation has been widely developed and studied by many researchers in medical AI domain~\cite{luu2021extending,peiris2022reciprocal}. For most of these methods,  U-Net~\cite{ronneberger2015u} and 3D U-Net~\cite{cciccek20163d} are the foundation architectures. These U-Shaped architectures enables local feature extraction while maintaining its contextual cues. However, what lacks in these CNN based methods are their inability to extract long range dependencies during training. Focusing more on local cues than global cues sometimes may lead to extract uncertain or imprecise information which degrades the segmentation performance and reliability.

Transformer architectures were originally proposed to address the above inductive bias issues. These transformer models are designed to extract long-range dependencies for sequence-to-sequence tasks~\cite{vaswani2017attention}.
Inspired by recent progress in Vision Transformers for Volumetric brain tumor segmentation~\cite{wang2021transbts,zhou2021nnformer,hatamizadeh2022unetr,peiris2021volumetric}, and it's unique abilities over Convolutional Neural Network (CNN) based models, we propose a U-shaped encoder-decoder neural network that adapts two way windowing approach during decoding for fine detail extraction. Vision Transformers~\cite{dosovitskiy2020image,cao2021swin,dong2021cswin} have shown ground breaking performance improvement in extracting long range dependencies by maintaining a flexible receptive fields. Also, the better robustness against data corruptions and occlusions~\cite{naseer2021intriguing}, shown in Transformer based deep learning models is unarguably the best thing that ever asked for in a neural network design. Considering these aspects of Transformer Network, in our proposed method we used two popular window based attention mechanisms namely, \underline{Cr}oss-Shaped window attention based \underline{Swin} Transformer block and Shifted window attention based \underline{Swin} Transformer block to constrcut a U-shaped \underline{V}olumetric \underline{T}ransformer (CR-Swin2-VT). In CR-Swin2-VT model, Swin Transformer blocks~\cite{cao2021swin} and CSWin Transformer blocks~\cite{dong2021cswin} are constructed parallel in the encoder side to capture voxel information precisely while only Swin Transformer blocks are used in the decoder side. In summary our major contributions are, \textbf{(1)}
 We propose a volumetric transformer architecture that can process medical scans as volumes entirety.
\textbf{(2)} We design encoding path with two window based attention mechanisms to capture local and global features of medical volumes.
\textbf{(3)} We conduct extensive experiments on FeTS Challenge 2022 dataset for brain tumor segmentation task.

\section{Method}
\label{sec:method}

\begin{figure*}[h]
\centering
\includegraphics[width=1\linewidth]{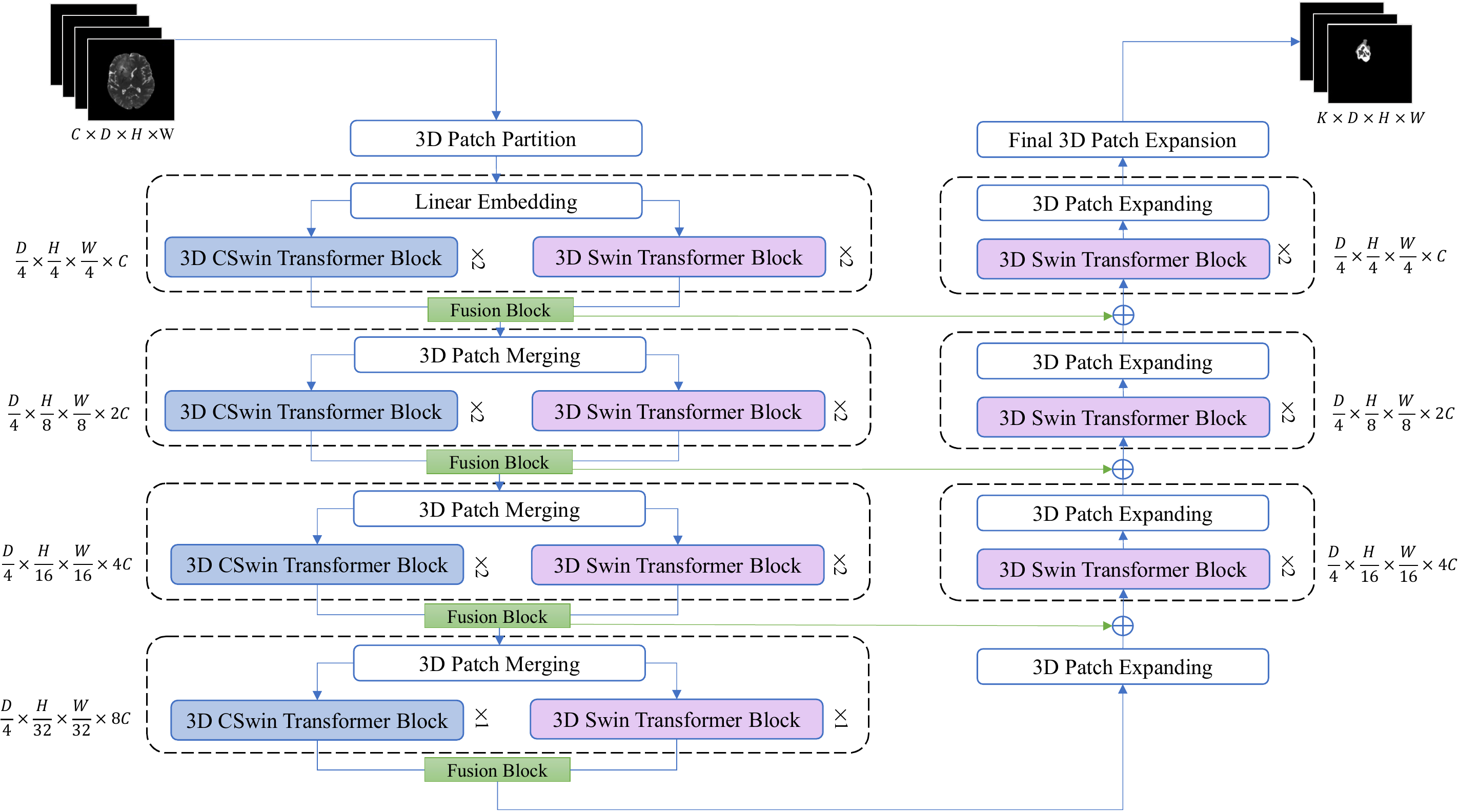} 
\caption{\textbf{Proposed CR-Swin2-VT Network Architecture.} }
\label{fig:architecture}
\end{figure*}

\begin{figure*}[h]
\scriptsize
\centering
\begin{tabular}{{c@{ } c@{ } c@{ }}}
     \multirow{3}{*}[0.5in]{{\includegraphics[width=0.52\linewidth]{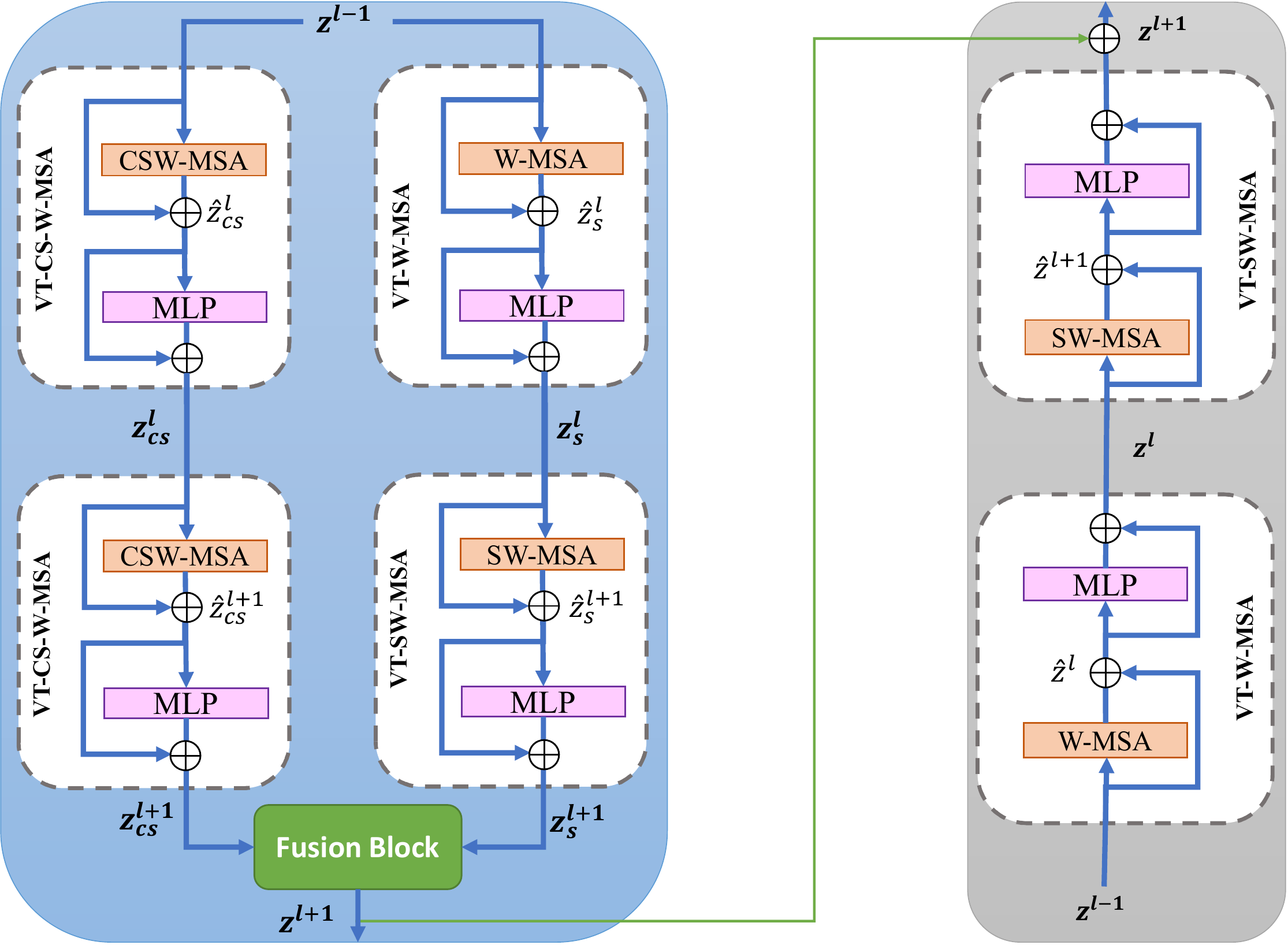}}} &
     \hspace{0.08cm} &
    {\includegraphics[width=0.35\linewidth]{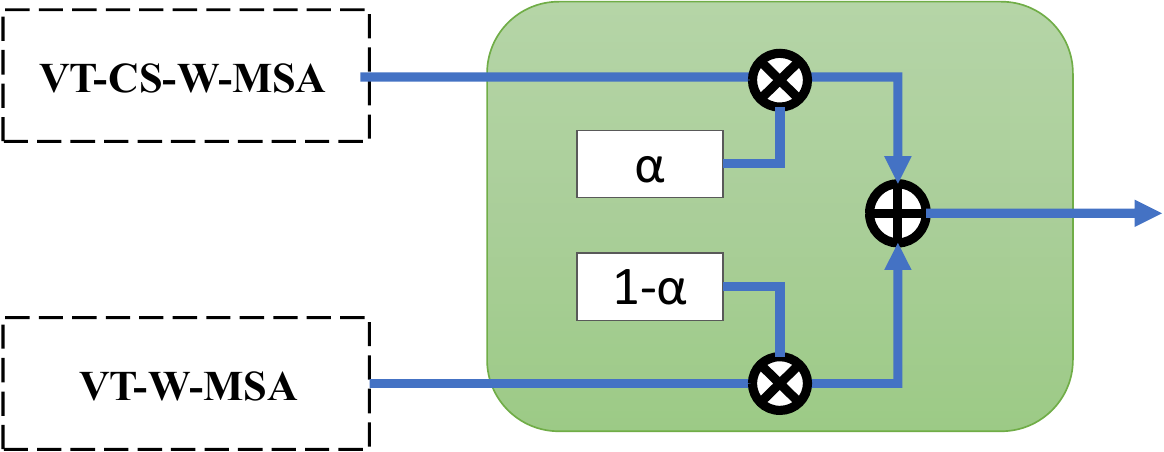}}\\
    & \hspace{0.08cm} & \footnotesize{(b)}\\
    & \hspace{0.08cm} & {\includegraphics[width=0.4\linewidth]{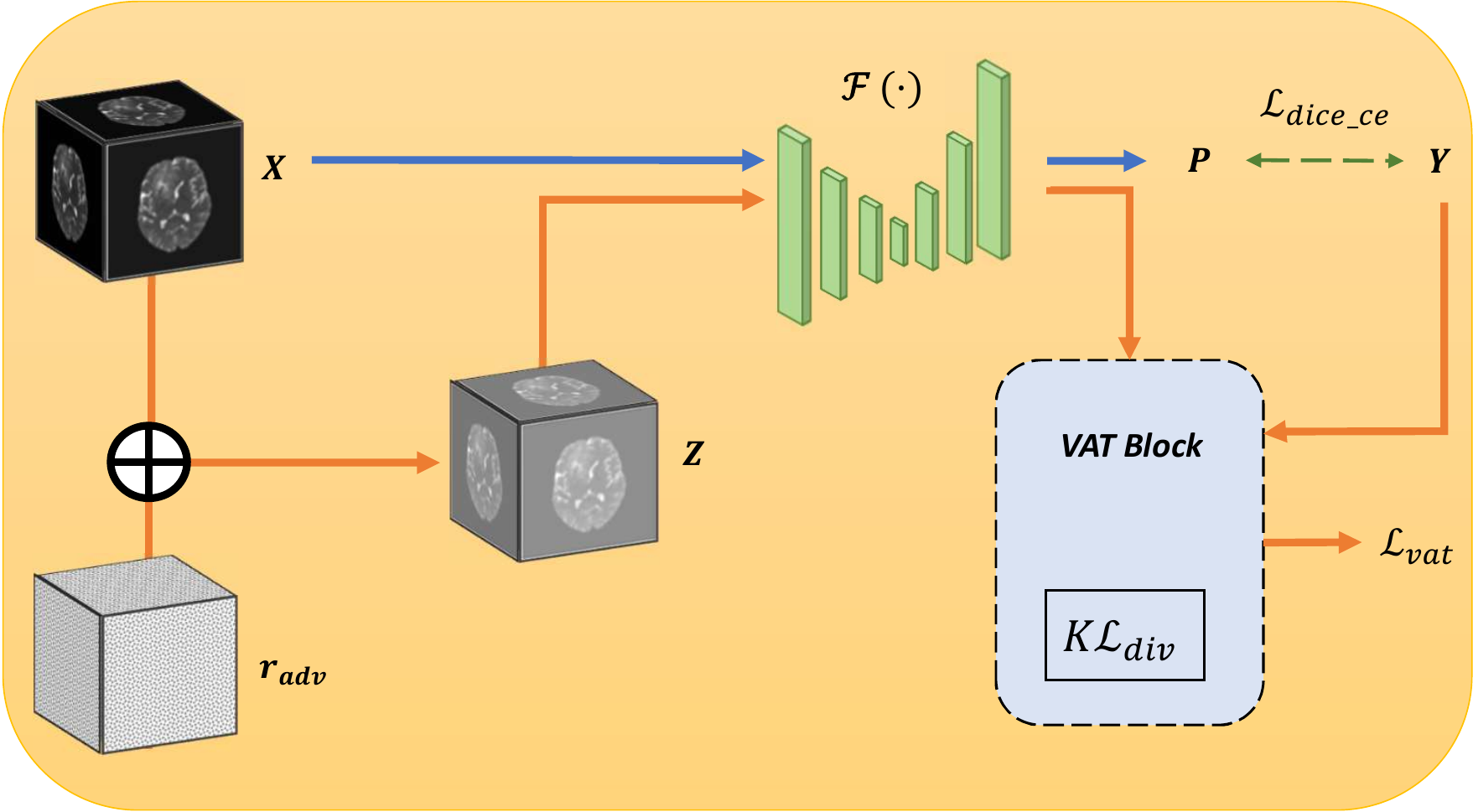}}\\
    \footnotesize{(a)} & \hspace{0.08cm} & \footnotesize{(c)}\\
  \end{tabular}
\caption{(a) Illustrates VT-UNet Architecture. Here, $k$ denotes the number of classes. (b) shows the structure of Fusion Block that aggregates two attention maps generate from CSwin Transformer block and Swin Transformer block in the decoder side of the network. (c) illustrates the CR-Swin2-VT model training process. Here, $X, Y, r_{adv}$ and $P$ are input data (original patient data), ground truth segmentation masks, perturbation added on input data and the prediction generated from segmentation network. }
\label{fig:modules}
\end{figure*}

We denote vectors and matrices in bold lower-case $\vec{x}$ and bold upper-case $\vec{X}$, respectively. 
Let $\mathds{X} = \{\vec{x}_1,\vec{x}_2,\cdots,\vec{x}_\tau\}, \vec{x}_i \in \mathbb{R}^C$ be a sequence representing the voxel patches of the medical volume (\eg, an MRI volume).
Here, each $\vec{x}_i$ is considered as a token. 
Following  previous work by Vaswani~\etal~\cite{vaswani2017attention}, we define  \underline{S}elf-\underline{A}ttention (SA) as:
\begin{equation}
    \text{SA}(\vec{Q}, \vec{K}, \vec{V}) = 
    \text{SoftMax}\Big(\vec{Q}\vec{K}^\top  \mathbin{/} \sqrt{C}\Big)\vec{V},
    \label{eqn:attn}
\end{equation}
where $\mathbb{R}^{\tau \times C_v} \ni \vec{V} = \vec{X}\vec{W}_\vec{V}$, by stacking tokens $\vec{x}_i$s into the rows of $\vec{X}$ (\ie, $\vec{X}=[\vec{x}_1|\vec{x}_2|\cdots\vec{x}_\tau]^\top$). 

This conventional SA mechanism is altered in many ways in recent transformer based studies, mainly by including typical positional encoding methods during SA calculation.  This allows adding back positional information back in action during model training. One of the most popular methods is \underline{R}elative \underline{P}ositional \underline{E}ncoding (RPE), which is used in celebrated work Swin Transformer~\cite{liu2021swin}. SA with RPE is defined as:
\begin{equation}
    \text{SA}(\vec{Q}, \vec{K}, \vec{V}) = 
    \text{SoftMax}\Big(\vec{Q}\vec{K}^\top  \mathbin{/} \sqrt{C}+\vec{B}\Big)\vec{V},
    \label{eqn:attn_rpe}
\end{equation}
where, $\mathbb{R}^{\tau \times \tau} \ni \vec{B}$ is trainable and acts as a relative positional bias across tokens in the volume with $\vec{V} = \vec{X}\vec{W}_\vec{V}$, $\vec{K} = \vec{X}\vec{W}_\vec{K}$, and $\vec{Q} = \vec{X}\vec{W}_\vec{Q}$. 

Dong~\etal proposed \underline{L}ocally-\underline{e}nhanced \underline{P}ositional \underline{E}ncoding (LePE), which adds the positional encoding as a parallel module to the SA operation~\cite{dong2021cswin}.
\begin{equation}
    \text{SA}(\vec{Q}, \vec{K}, \vec{V}) = 
    \text{SoftMax}\Big(\vec{Q}\vec{K}^\top  \mathbin{/} \sqrt{C}\Big)\vec{V} +\vec{L},
    \label{eqn:attn_lepe}
\end{equation}
where, $\mathbb{R}^{\tau \times C_v} \ni \vec{L}$ is trainable and acts as a locally-enhanced positional bias operates on projected \emph{Values}, $\vec{V} = \vec{X}\vec{W}_\vec{V}$ in each CSwin Transformer block and this mechanism can enforce stronger local inductive bias. Therefore, our proposed CR-Swin2-VT architecture adapts two attention based windowing mechanisms with distinct positional encoding methods to extract strong long-range features accurately during contracting path of encoder-decoder design.

\subsection{Overall Architecture}
The overall architecture of CR-Swin2-VT is illustrated in~\cref{fig:architecture}. The input to CR-Swin2-VT model is a 3D volume of size $D \times H \times W \times C$. The output is a $D \times H \times W \times K$ dimensional volume. Here, $k$ denotes the number of classes. Similar to~\cite{peiris2021volumetric}, the proposed model comprises of CR-Swin2-VT Encoder, CR-Swin2-VT Bottleneck and CR-Swin2-VT Decoder.

\subsubsection{CR-Swin2-VT Encoder.}
CR-Swin2-VT Encoder consists of 3D Patch Partitioning layer combined with Linear embedding layer, 3D CSwin Transformer blocks, 3D Swin Transformer blocks, Fusion block and 3D Patch merging layer. During 3D patch partitioning, the input medical volumes (e.g., T1, T1CE, Flair, T2 MRI sequences) are split into non-overlapping voxel/3D patches and feed to the next linear embedded layer as a set of tokens. Here we used a partitioning kernel $P \times M \times M$ where $P=2$ and $M=4$, which results in $2 \times 4 \times 4$ patch partitioning kernel. During linear embedding the resultant tokens are mapped into a $C$ dimensional vector (Embedded Dimensions). In our experiments, we set $C=48$. These tokens are then passed through two successive 3D Swin Transformer blocks~\cite{liu2021swin}(\textbf{VT-W-MSA-Blk}), which are comprised of \textbf{(1)} Window Multi Head SA (W-MSA), \textbf{(2)} Shifted Window Multi Head SA (SW-MSA) and two successive CSwin Transformer blocks~\cite{dong2021cswin}(\textbf{VT-CS-W-MSA-Blk}) which are consisted of Cross shaped Window SA (CSW-MSA). 

\paragraph{\textbf{VT-W-MSA-Blk.}}
During W-MSA operation, the volume is evenly split into smaller non-overlapping windows and attention is calculated for those windows. In order to extract long range dependencies, shifted window approach is used during SW-MSA operation~\cite{liu2021swin}. 
Therefore, VT-W-MSA Block's functionality is defined as:
\begin{align}
    &{{\hat{\bf{z}}}^{l}_{s}} = \text{W-MSA}\left( {\text{LN}\left( {{{\bf{z}}^{l - 1}}} \right)} \right) + {\bf{z}}^{l - 1},\nonumber \qquad 
    &{{\hat{\bf{z}}}^{l+1}_{s}} = \text{SW-MSA}\left( {\text{LN}\left( {{{\bf{z}}^{l}_{s}}} \right)} \right) + {\bf{z}}^{l}_{s}, \nonumber\\
    &{{\bf{z}}^{l}_{s}} = \text{MLP}\left( {\text{LN}\left( {{{\hat{\bf{z}}}^{l}_{s}}} \right)} \right) + {{\hat{\bf{z}}}^{l}_{s}}, \qquad 
    &{{\bf{z}}^{l+1}_{s}} = \text{MLP}\left( {\text{LN}\left( {{{\hat{\bf{z}}}^{l+1}_{s}}} \right)} \right) + {{\hat{\bf{z}}}^{l+1}_{s}}, 
    \label{eqn:vt_swin_block}
\end{align}
where ${\hat{\bf{z}}}^l_s$ and ${\bf{z}}^l_s$ denote the output features of the W-MSA module and the Multi Layer Perceptron (MLP) module for block $l$, respectively. A
Layer Normalization (LN) is applied before every MSA and MLP, and a residual connection is applied after each module. 

\paragraph{\textbf{VT-CS-W-MSA-Blk.}}
CSW-MSA operation consisted of calculating SA in horizontal and vertical stripes in parallel that form a cross-shaped window~\cite{dong2021cswin}. The VT-CS-W-MSA Block's functionality is defined as:
\begin{align}
    &{{\hat{\bf{z}}}^{l}_{cs}} = \text{CSW-MSA}\left( {\text{LN}\left( {{{\bf{z}}^{l - 1}}} \right)} \right) + {\bf{z}}^{l - 1},\nonumber \qquad 
    &{{\hat{\bf{z}}}^{l+1}_{cs}} = \text{CSW-MSA}\left( {\text{LN}\left( {{{\bf{z}}^{l}_{cs}}} \right)} \right) + {\bf{z}}^{l}_{cs}, \nonumber\\
    &{{\bf{z}}^{l}_{cs}} = \text{MLP}\left( {\text{LN}\left( {{{\hat{\bf{z}}}^{l}_{cs}}} \right)} \right) + {{\hat{\bf{z}}}^{l}}, \qquad 
    &{{\bf{z}}^{l+1}_{cs}} = \text{MLP}\left( {\text{LN}\left( {{{\hat{\bf{z}}}^{l+1}_{cs}}} \right)} \right) + {{\hat{\bf{z}}}^{l+1}_{cs}}, 
    \label{eqn:vt_cswin_block}
\end{align}
where ${\hat{\bf{z}}}^l_{cs}$ and ${\bf{z}}^l_{cs}$ denote the output features of the CSW-MSA module and the Multi Layer Perceptron (MLP) module for block $l$, respectively. A Layer Normalization (LN) is applied before every MSA and MLP, and a residual connection is applied after each module.

\paragraph{\textbf{Fusion Block.}} The output tokens generated from each VT-W-MSA-Blk and VT-CS-W-MSA-Blk are then aggregated using fusion function $\mathcal{F}(\cdot)$ which gives $\vec{z}^{l+1}$. $\mathcal{F}(\cdot)$ is defined as:
\begin{align}
    \mathcal{F}(\vec{z}^{l+1}_s, \vec{z}^{l+1}_{cs}) = \alpha~ {{\vec{z}^{l+1}_s}} + (1 - \alpha)~{{\vec{z}^{l+1}_{cs}}},
    \label{eqn:fusion_function}
\end{align}
where, we use a linear combination with $\alpha =  0.5$. The aggregated output produced from fusion block is then passed into 3D patch merging layer to generate feature hierarchies in the encoder of CR-Swin2-VT. 

\subsubsection{Bottleneck.} The bottleneck consisted of one successive block from VT-W-MSA-Blk and VT-CS-W-MSA-Blk together with 3D Patch Expanding layer.

\subsubsection{CR-Swin2-VT Decoder.}
The CR-Swin2-VT decoder starts with successive VT-W-MSA-Blks together with 3D patch expanding layers and a classifier at the end to generate final volumetric segmentation masks. 

\subsection{Training Objective}
The CR-Swin2-VT model's objective is to segment volumetric medical images and it's model learning process is morel image segmentation and the training process is more deeply shared across the loss functionthat we used.

\subsubsection{Loss Function.}
Let  $\mathcal{X} = \{(\vec{X}_1,\vec{Y}_1), \cdots , (\vec{X}_n,\vec{Y}_n)\}$ denote the labeled data from $n$ patients, where each pair $(\vec{X}_i,\vec{Y}_i)$ has an image $\vec{X}_i \in \R^{D \times H \times W \times C}$ and its associated ground-truth mask $\vec{Y_i} \in \R^{D \times H \times W \times K}$. To train CR-Swin2-VT, we jointly minimize the Dice Loss (DL), Cross Entropy (CE) loss and VAT loss. The three loss terms are modified and computed in a voxel-wise manner. The DL is defined as:
\begin{align}
    \Loss_{\mathrm{dl}}(\theta;\mathcal{X}) \hspace{-0.1ex} &= - \mathbb{E}_{(\vec{X}, \vec{Y}) \sim \mathcal{X}} \Bigg[ \frac{ 2 
     \big \langle \vec{Y} \hspace{-0.1ex}~,~ \hspace{-0.1ex} \mathcal{H}(\vec{X}) 
     \hspace{-0.2ex} \big \rangle}
     {\big\|\vec{Y}\big\|_1 + \big\|\mathcal{H}(\vec{X})\big\|_1} \Bigg],
     \label{eqn:dice_loss}
\end{align}
where $\mathcal{H}(\cdot)$ and $\theta$ denote the transformer model and the model parameters, respectively. 
The CE loss is defined as:
\begin{align}
    \Loss_{\mathrm{ce}}(\theta;\mathcal{X}) \hspace{-0.1ex} &= \mathbb{E}_{(\vec{X}, \vec{Y}) \sim \mathcal{X}} \Big[ - \vec{Y} \log \mathcal{H}(\vec{X}) \Big],
     \label{eqn:ce_loss}
\end{align}

During CR-Swin2-VT model training we make use of VAT to update the model by the weighted sum of the gradient and considered the loss introduced during VAT as a regularization term for out full objective function. Inspired by the VAT method by Takeru \etal~\cite{miyato2018virtual}, $\Loss_{vat}$ is calculated as Kullback-Leibler (KL) divergence loss which measures the divergence between ground truth distribution and perturbed prediction distribution.  The VAT block improves the CR-Swin2-VT model's robustness against adversarial samples that violates the virtual adversarial direction. Therefore, the VAT loss term is defined as a divergence based Local Distributional Smoothness (LDS):
\begin{align}
    \label{eqn:vat_loss}
    \Loss_{\mathrm{vat}}(\theta_G;\mathcal{X}; r_{adv}) \hspace{-0.1ex} &= \mathbb{E}_{(\vec{X}, \vec{Y}) \sim \mathcal{X}} \Bigg[ \mathcal{D}_{KL}(\vec{Y} \big\| \mathcal{F}(\theta_G, \vec{X} + r_{adv}))\Bigg].
\end{align}
where $\lambda$ is a hyper-parameter which controls the contribution of VAT loss term. 
Therefore, the full objective function is:
\begin{align}
    \Loss_{\mathrm{seg}}(\theta;\mathcal{X}) \hspace{-0.1ex} &= \Loss_{\mathrm{dl}}(\theta;\mathcal{X}) + \Loss_{\mathrm{ce}}(\theta;\mathcal{X}) + \lambda  \Loss_{vat}(\theta; \mathcal{X}; r_{adv})
     \label{eqn:seg_loss}
\end{align}
\section{Experiments}
\label{sec:experiments}

\subsubsection{Dataset.}
We use mpMRI from the FeTS Challenge 2022~\cite{pati2021federated,reina2021openfl,baid2021rsna,sheller2020federated} for CR-Swin2-VT model training and evaluation.
The training dataset has 1251 MR volumes of shape $240 \times 240 \times 155$ from four MRI sequences, that are conventionally used for giloma detection: T1 weighted sequence (T1), T1-weighted contrast enhanced sequence using gadolinium contrast agents (T1Gd) (T1CE), T2 weighted sequence (T2), and Fluid attenuated inversion recovery (FLAIR) sequence. 
These sequences are then used to identify, four distinct tumor sub-regions as: The Enhancing Tumor (ET) which corresponds to area of relative hyper-intensity in the T1CE with respect to the T1 sequence, Non Enhancing Tumor (NET), Necrotic Tumor (NCR) which are both hypo-intense in T1-Gd when compared to T1, Peritumoral Edema (ED) which is hyper-intense in FLAIR sequence.
These almost homogeneous sub-regions are then converted into three semantically meaningful tumor classes as: Enhancing Tumor (ET), addition of ET, NET and NCR represents the Tumor Core (TC) region and addition of ED to TC represents the Whole Tumor (WT).

\subsubsection{Image Pre-processing.}
Intensities of MRI volumes are inconsistent due to various factors such as motions of patients during the examination, different manufacturers of acquisition devices, sequences and parameters used during image acquisition. To standardize all volumes, min-max scaling was performed followed by clipping intensity values. Images were then cropped to a fixed patch size of $128 \times 128 \times 128$ by removing unnecessary background pixels. 

\subsubsection{Implementation Details.}
The proposed CR-Swin2-VT model is implemented in PyTorch with a single Nvidia RTX 3090 GPU with 24GB. The weights of Swin-T~\cite{liu2021swin} pre-trained on ImageNet-22K are used to initialize the model.  
To train CR-Swin2-VT we use Adam optimizer with the learning rate of 1e-04 with batch size of 1, 1000 epochs and ploy decay for learning rate scheduling. We split the original FeTS 2022 training dataset into training set (80\%) and validation set (20\%). Therefore, 1000 MR volumes are used to train the model while 251 MR volumes were used as validation set to evaluate model's performance on unseen patient data during training. The best performing model for validation set is saved as the best model for official validation and testing phase evaluation.
The FeTS 2022 validation dataset contains 219 MR volumes and synapse portal conducts the evaluation. In the inference phase, the original volume re-scaled using min-max scaling and feed forward through the CR-Swin2-VT model. During inference, we use sliding window approach with the patch size of $128 \times 128 \times 128$.

\subsubsection{Evaluation Matrice.s}
Segmentation accuracy of three classes (\ie, ET, TC and WT) are evaluated during training and inference. Both qualitative and quantitative analysis is performed to evaluate the model accuracy. The proposed CR-Swin2-VT model is evaluated using four matrices (1) Dice S$\o$rensen coefficient (DSC), (2) Hausdorff Distance, (3) Sensitivity and (4) Specificity. 

\begin{table*}[!htb]
\small
\centering
\caption{Validation Phase Results.}
\begin{tabular}{l | c | c | c | c | c  }
\hline
Method & Class & ~Hausdorff Distance~ & ~Dice Score~ & ~Sensitivity~ & ~Specificity~\\
\hline
\hline
\multirow{3}{*}{\makecell{CNN VAT based \\ Method~\cite{peiris2022reciprocal}}} & ET & 21.83 &	81.398 & 83.40 &	99.97 \\
& TC & 8.56 &	85.39 &	85.07 &	99.98 \\
& WT & 5.37 &	90.77 & 92.09 &	99.91 \\

\hline

\multirow{3}{*}{CR-Swin2-VT} & ET & 14.81 &	81.71 & 82.38 &	99.98 \\
& TC & 11.19 &	85.40 &	84.33 &	99.98 \\
& WT & 3.93 &	91.38 & 91.23 &	99.93 \\
\hline
\end{tabular}
\label{fig:quantitative}
\end{table*}

\subsubsection{Experimental Results.}
\begin{figure*}[!htb]
\scriptsize
\tabcolsep=0.04cm
\centering
  \begin{tabular}{{c@{ }}}
    {\includegraphics[width=0.60\linewidth]{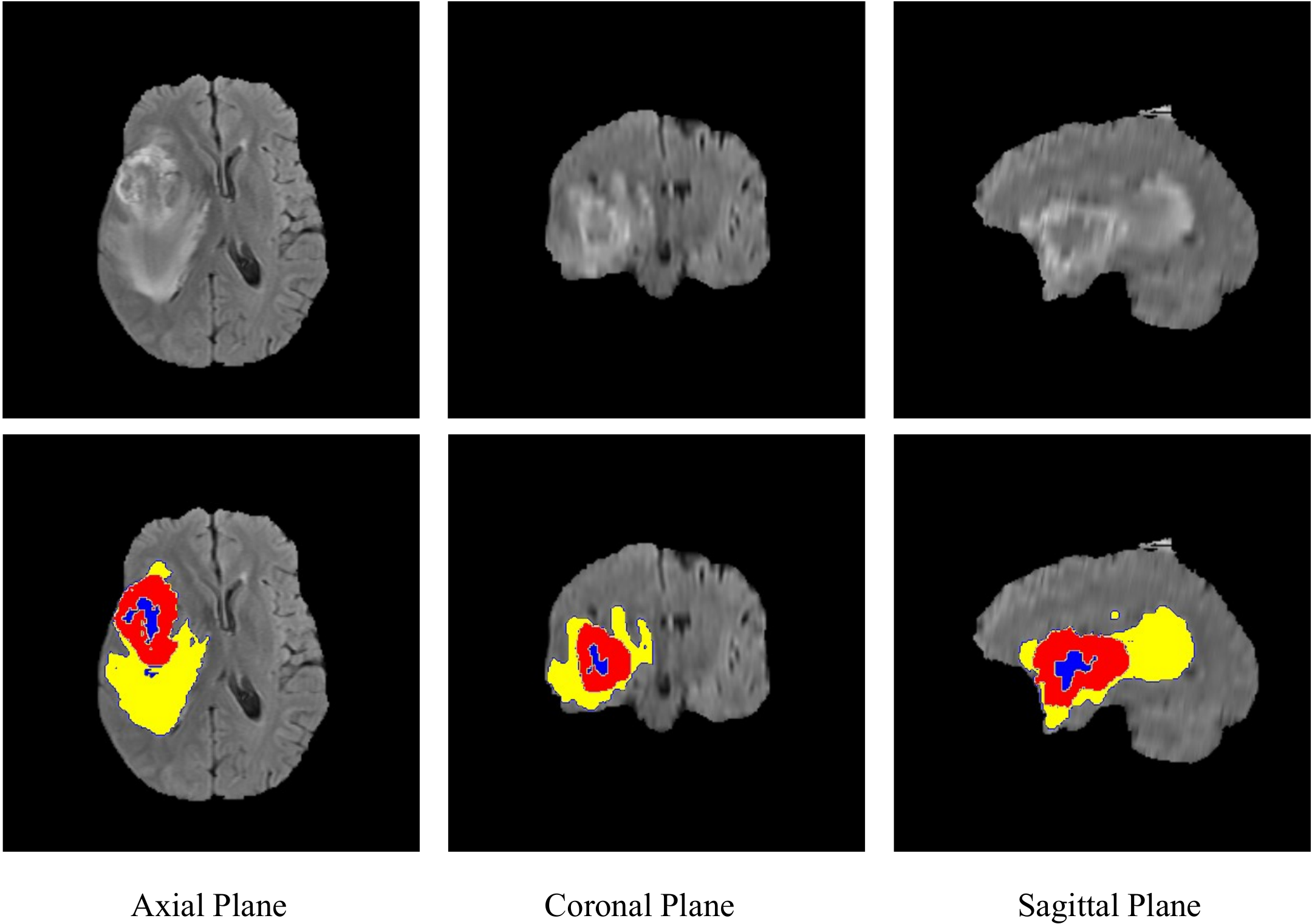}}
  \end{tabular}
    \caption{Validation Phase Results for the Sample FeTS2022$\_$01779. Here, yellow, red and blue represents the peritumoral edema (ED), Enhancing Tumor (ET) and non enhancing tumor/necrotic tumor (NET/NCR), respectively. (Dice (ET) = 96.37, Dice (TC) = 97.21, Dice (WT) = 97.12)}
    \label{fig:qualitative}
\end{figure*}

\begin{figure*}[!htb]
\scriptsize
\tabcolsep=0.04cm
\centering
  \begin{tabular}{{c@{ }}}
    {\includegraphics[width=1\linewidth]{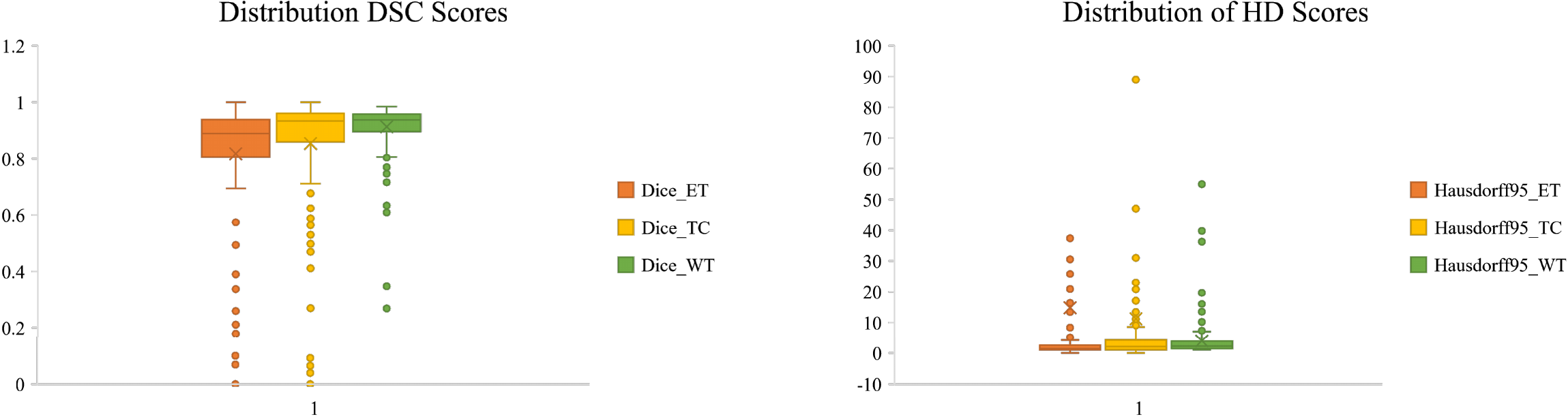}}\\
  \end{tabular}
    \caption{The box and whisker plots of the distribution of the segmentation metrics for Validation Phase Results. The box-plot shows the minimum, lower quartile, median, upper quartile and maximum for each tumor class. Outliers are shown away from lower quartile.}
    \label{fig:box_plot}
\end{figure*}

The quantitative and qualitative results of online validation phase evaluation for the proposed approach is shown in Table~\ref{fig:quantitative} Fig.~\ref{fig:box_plot} and Fig.~\ref{fig:qualitative}. 

\section{Conclusion}
\label{sec:conclusion}
We proposed a Transformer-based method that has adapted two windowing strategies in encoder to extract the long-range dependencies both within and across different modalities of mpMRI.  We validated our method on brain tumor segmentation using FeTS Challenge 2022 dataset and results demonstrate the effectiveness of the proposed method.

\bibliographystyle{splncs04}
\bibliography{references}

\end{document}